# *Mechanistic Models in Computational Social Science*

Petter Holme[1*], Fredrik Liljeros[2]

[1] Department of Energy Science, Sungkyunkwan University, Suwon 440-746, Korea

[2] Department of Sociology, Stockholm University, 10961 Stockholm, Sweden

[*] Correspondence: Petter Holme, holme@skku.edu

## Abstract

Quantitative social science is not only about regression analysis or, in general, data inference. Computer simulations of social mechanisms have a 60-year long history. They have been used for many different purposes—to test scenarios, test the consistency of descriptive theories (proof-of-concept models), explore emergent phenomena, forecast, etc. In this essay, we sketch these historical developments, the role of mechanistic models in the social sciences, and the influences from the natural and formal sciences. We argue that mechanistic computational models form a common ground for social and natural sciences and look forward to possible future information flow across the social-natural divide.

## Background

In mainstream empirical social science, a result of a study often consists of two conclusions. First, there is a statistically significant correlation between a variable describing a social phenomenon and a variable thought to explain it. Second, the correlations with other, more fundamental, or trivial variables (called *control*, or *confounding*, variables) are weaker. In recent years, there has been a trend to criticize this approach for putting too little emphasis on the mechanisms behind the correlations [Woodward 2014, Salmon 1998, Hedström Ylikoski 2010]. Authors often argue that regression analysis (and the linear, additive models they assume) cannot be causal explanations of an open system, as usually studied in social science. The main reason is that, in an empirical study, there is no way of isolating all plausible mechanisms [Sayer 2000]. Sometimes authors point to natural science as a role model in the quest for mechanistic models. This is somewhat ironic since many natural sciences, most notably physics, traditionally emphasize the unification of theories and the reduction of hypotheses [Woodward 2014]. In other words, they are striving to show that two theories could be more simply described as different aspects of a single, unified theory. Rather than being imported from natural or formal sciences, mechanistic modeling has evolved in parallel in social science. Maybe the most clean-cut forms of mechanistic models are those used in computer simulations. Their past, present, and future, and the flow of information regarding them across disciplines, are the themes of this paper. Before proceeding, other authors would probably spend considerable amounts of ink to define and discuss central concepts—in our case, "mechanism" and "causal." We think their everyday usage in both natural and social sciences is sufficiently precise for our purpose and recommend Hedström and Ylikoski [2010] to readers with a particular interest in details.

In practice, establishing the mechanisms behind a social phenomenon takes much more than simulating a model. Mechanistic models can serve several different purposes en route to establishing a mechanistic explanation. We will distinguish between *proof-of-concept modeling*, *the discovery of hypotheses*, and *scenario testing* (described in detail below). There are, of course, other ways, perhaps also better, to characterize mechanistic models. These categories are not strict either—they could be overlapping concerning a specific model. Nevertheless, we think they serve a point in our discussion and that they are reasonably well defined.

The idea of proof-of-concept modeling is to test the consistency of a verbal description, or cartoon diagram, of a phenomenon [Servedio & al. 2014]. It is generally hard to make an accurate verbal explanation, especially if it involves connecting different levels of abstraction, such as going from a microscopic to a macroscopic description. A com-mon mistake is to neglect implicit assumptions, some that may even be the convention of a field. With the support of such proof-of-concept models, a verbal argument becomes much more substantial. Then one has at least firmly established that the constituents of the theory are sufficient to explain the phenomenon. The individual-based simulations of the Anasazi people (inhabiting parts of the American West millennia ago) by Joshua Epstein, Robert Axtell, and colleagues [Epstein Axtell 1996] are blueprints of proof-of-concept modeling. In these simulations, the authors combined a multitude of conditions along with anthropological theories to show that they could generate outcomes similar to the archaeological records.

The most common use of mechanistic models is our second category—to explore the possible outcomes of a specific situation and to generate hypotheses. We will see many examples of that in our essay. As a first example, consider Robert Axelrod's computer tournaments to find optimal strategies for the iterated prisoner's dilemma [Axelrod 1984]. The prisoner's dilemma captures a situation where an individual can choose whether or not to cooperate with another. If one knows that the encounter is the last one, the rational choice is always not to cooperate. However, if the situation could be repeated an unknown number of times, it might be better to cooperate. To figure out how to cope with this situation, Axelrod invited researchers to submit strategies to a round-robin tournament. The winning strategy ("tit-for-tat") was to start cooperating and then do whatever your opponent did the previous step. From this result, Axelrod could make the hypothesis that a tit-for-tat-like behavior is typical among both people and animals, either because they often face a prisoner's dilemma or at that such situations, once they face them, tend to be important.

Mechanistic models forecasting social systems are less frequent than our previous two classes. One reason is probably that forecasting open systems is difficult (sometimes probably even impossible) [Sayer 2000]; another that non-mechanistic methods (machine learning, statistical models, etc.) are better for this purpose. A model without any predictive power whatsoever is, of course, not a model at all, and under some conditions, all mechanistic models can be used in forecasting or (perhaps more accurately) scenario testing. One notable example is the "World3" simulation popularized by the Club of Rome 1972 book *The Limits to Growth* [Meadows & al. 1972], where an exponentially growing artificial population faced a world of limited resources. Maybe a sign of the time, since several papers from the early 1970s called for "whole Earth simulations" [Patterson 1970, Rau 1970]. Echoes of this movement were heard recently with the proposal of a "Living Earth Simulator" [Paolucci & al. 2012].

In this essay, we will explore mechanistic models as scientific explanations in the social sciences. We will give an overview of the development of computer simulations of mechanistic models (primarily in the social sciences, but also mentioning relevant developments in the natural sciences), and finally discuss if and how mechanistic models can be a common ground for cross-disciplinary research between the natural and social sciences. We do not address data-driven science in the interface of the natural or social, nor do we try to give a comprehensive survey of mechanistic models in the social sciences. We address anyone interested in using simulation methods familiar to theoretical natural scientists to advance the social sciences.

## Influence from the natural and formal sciences

As we will see below, the development and use of computer simulations to understand social mechanisms have happened on relatively equal terms as in the natural and formal sciences. It will, however, be helpful for the subsequent discussion to sketch the critical developments of computer simulations as mechanistic models in the natural sciences. This is, of course, a topic that would need several book volumes for a comprehensive coverage—we will just mention what we regard as the most important breakthroughs.

### *The military origins*

Like in social science, simulation in natural science has many of its roots in the military from around the Second World War. The second major project running on the first programmable computer, ENIAC, started in April 1947. The topic, the flow of neutrons in an incipient explosion of a thermonuclear weapon [Haigh & al. 2014], is perhaps of little interest today, but the basic method has never run out of fashion—it was the first computer program using (pseudo) random numbers, and hence an ancestor of most modern computer simulations. Exactly who invented this method, codenamed Monte

Carlo, is somewhat obscure, but it is clear it came out of the development of the hydrogen bomb right after the war. The participants came from the (then recently finished) Manhattan project. Nicholas Metropolis, Stanislaw Ulam, and John von Neumann are perhaps most well known, but also Klara von Neumann, John's wife [Haigh & al. 2014]. It was not only the first program to use random numbers; it was also the first modern program in the sense that it had function calls and had to be fed into the computer along with the input. As a curiosity, the random number generator in this program worked by squaring eight-digit numbers and using the mid-eight digits as output and seed to the next iteration. Far from having the complexity of modern pseudo-random number generator (read Mersenne Twister [Matsumoto Nishimura 1998]), it gives random numbers of (at least in the authors' opinion) surprisingly good statistical quality.

The first Monte Carlo simulation was not an outright success as a contribution to the nuclear weapons program. Nevertheless, the idea of using random numbers in simulations has not fallen out of fashion ever since, and the Monte Carlo method (nowadays referring to any computational model based on random numbers) has become a mainstay of numerical methods. Another significant step for the natural sciences, especially chemistry and statistical physics, by the Los Alamos group was the Metropolis-Hastings algorithm—a method to sample configurations of particles, atoms, or molecules according to the Boltzmann distribution (connecting the probability of a configuration and its energy). The radical invention was to choose configurations with a probability proportional to the Boltzmann distribution and weighing them equally, rather than choosing configurations randomly and weighing them by the probability given by the Boltzmann distribution [Metropolis & al. 1953]. Hastings name was added to credit his extension of the algorithm to general distributions [Hastings 1970]. Today, this algorithm is an indispensable simulation technique to generate the probability distributions of the state of a system both in natural and social sciences (usually called Markov Chain Monte Carlo, MCMC).

The Monte Carlo project and the MCMC method did not immediately lead to fundamental advances in science itself. On the other hand, deterministic computational methods did, and (not surprisingly) post-Manhattan-project researchers were involved. Enrico Fermi, John Pasta, and Stanislaw Ulam (and, like the Monte Carlo project, with uncredited help by a female researcher, Mary Tsingou [Dauxois 2008]) studied vibrations of a one-dimensional string with nonlinear corrections to Hooke's law (which states that the force needed to extend a spring a certain distance is proportional to the distance). They expected to see the nonlinearity transferring energy from one vibrational mode (like the periodic solution of the linear problem) to all other modes (i.e., thermal fluctuations) according to the equipartition theorem [Landau Lifshitz 1980]. Instead of such a "thermalization" process, they observed the transition to a complex, quasi-periodic state [Fermi & al. 1955] that never lost its memory of the initial condition. The FPU paradox was the starting point of a scientific theme called nonlinear science that also, as we will see, has left a lasting imprint on social science.

*Complexity theory*

Nonlinear science strongly overlaps with chaos theory, another set of ideas from natural sciences that influenced social science. Chaos is summarized in the vernacular by the "butterfly effect"—a small change (the flapping of a butterfly's wings) could lead to a big difference (a storm) later. One crucial early contribution came from Edward Lorentz's computational solutions of equations describing atmospheric convection. He observed that a slight change in the initial condition could send the equations off into entirely different trajectories [Lorentz 1963]. Just like for the FPU paradox, the role of the computational method in chaos theory has primarily been to discover hypotheses that later have been corroborated by analytical studies. This line of research has not been directly aimed at discovering new mechanisms; still, ideas and concepts from chaos theory have also reached social sciences [Kiel Elliott 1996].

Another natural science development largely fueled by computer simulations, which has influenced social sciences, is fractals. Fractals are mathematical objects that embody self-similarity—a river can branch into tributaries, that branch into smaller tributaries, and so on until the biggest rivers are reduced to the tiniest creeks [Mandelbrot 1983]. At all scales, the branching looks the same. Fractals provide an analysis tool—the fractal dimension—that can characterize self-similar objects. Many socioeconomic systems are self-similar—financial time series [Mantegna Stanley 1999], the movement of people [Brockmann & al. 2006], the fluctuations in the size of organizations [Mondani & al. 2014], etc. Quite frequently, however, authors have not accompanied their measurement of a fractal dimension with a mechanistic explanation of it, which is perhaps why fractals have fallen out of fashion lately.

Fractals are closely related to power-law probability distributions, i.e., the probability of an observable $x$ being proportional to $x^{-a}$, $a > 0$. Power-laws are the only self-similar (or "scale-free") real-to-real functions in the sense that, if, e.g., the wealth distribution of a population is a power law, then a statement like "there are twice as many people with a wealth of $10x$ than $15x$" is true, no matter if $x$ is dollars, euros, yen or kronor [Newman 2005]. The theories for such power-law phenomena date back to Pareto's lectures on economics published in 1896 [Pareto 1896]. Fractals and power laws are also connected to phase transitions in physics—an idea popularized in Hermann Haken's book Synergetics [Haken 1982].

The next step in our discussion is the studies of artificial life. The central question in this line of research is to mechanistically recreate the fundamental properties of a living system, including self-replication, adaptability, robustness, and evolution [Langton 1998]. The origins of artificial life can be traced to John von Neumann's self-replicating cellular automata. These are configurations of discrete variables confined to an underlying square grid that, following a distinct set of rules, can reproduce, live and die [von Neumann 1966]. The field of artificial life later developed in different directions, both toward the more abstract study of cellular automata and to more biology-related questions [Langton 1998]. It is also strongly linked to the study of adaptive systems (systems able to respond to environmental changes) [Miller Page 2009] and has a few recurring ideas related to social phenomena. The first idea is that simple rules can create complex behavior. The best-known model illustrating this is perhaps Conway's game of life. This is a cellular automaton with the same objectives as von Neuman but with fewer and simpler rules [Langton 1998]. The second idea (maybe not discovered by the field of artificial life, but at least popularized) is that of emergence. This refers to the properties of a system, as a whole, coming from the interaction of a large number of individual subunits. A textbook example is that of murmurations of birds ( flocks of hundreds of thousands of e.g., starlings). These can exhibit an undulating motion, fluctuating in density, that in no way could be anticipated from the movement of an individual. Another feature of emergence, exemplified by bird flocks, is decentralization—there is no leader bird. These topics are common to many social science disciplines (emergence is similar to the micro-to-macro-transition in sociology and economics). These theories have spawned their own modeling paradigm—agent-based models [Šalamon 2011, Epstein 2006, Carley Wallace 2001, Hedström Manzo 2015]—similar to what was simply called "simulation" in early computational social science. One first sets up rules for how units (agents) interact with each other and their surroundings. Then one simulates many of them together (typically on a two-dimensional grid) and lets them interact. We note that the concept of emergence has also been influential to cognitive, and subsequently behavioral, science. The idea of cognitive processes being emergent properties of neural networks—connectionism [Dawson 2008]—is nowadays fundamental to our understanding of computational processes in nature [Flake 1998].

In the 1980s, artificial life, adaptive systems, fractals, and chaos were grouped together under the umbrella term complexity science [Mitchell 2011]. This was, in many ways, a social movement gathering researchers of relatively marginalized research topics (the Santa Fe Institute and other similar centers acted as hubs for this development). Many of the themes within complexity science could probably just as well be categorized as mutually independent fields. This is perhaps best illustrated in that there is no commonly accepted definition of "complexity." Instead, there are several common, occasionally (but not always) connected themes (like the above-mentioned emergence, decentralized organization, fractals, chaos, etc.) that together define the field. On the other hand, complexity scientists have a common goal to find general, organizational principles that are not limited to one scientific field. In spirit, this dates back to, at least, von Bertalanffy's general systems theory [von Bertalanffy 1968]. The diversity of ideas and applications has not necessarily been a problem for complexity science; on the contrary, it has encouraged many scientists of different backgrounds (including the authors of this paper) to collaborate, despite the transdisciplinary language barriers.

*Game theory*

Game theory is a mathematical modeling framework for situations where the state of an individual is jointly determined by the individual's own decisions and the decisions of others (who all, typically, strive to maximize their own benefit) [Hofbauer Sigmund 1998]. Vaccination against infectious diseases is a typical example. If everyone else were vaccinated, the rational choice would be not to get vaccinated. The disease could anyway not spread in the population, whether or not you are vaccinated. Moreover, vaccines can, after all, have side effects, and injections are uncomfortable. If nobody were vaccinated, and the chance of getting the disease times the gravity of the consequences outweighs the inconveniences mentioned above, then it would be rational to get vaccinated. This situation could, mathematically, be phrased as a minority game [Challet & al. 2005]. The emergent solution for a population of rational, well-informed, and selfish individuals is that a fraction of the agents would get vaccinated and another fraction not. At the time of writing, this example is the background to a controversy where people getting vaccinated see people resisting vaccination as irresponsible to society [Honigsbaum 2015].

Game theory has been a powerful undercurrent in economics and population biology. We note that a special feature of game theory, compared to similarly interdisciplinary theories, is that the various fields using it seem well informed about the other fields' progress, and not so many concepts have been reinvented. Game theory itself is not a framework for mechanistic models, and especially in population biology (where an individual usually represents a species or a sub-population), it is not its primary purpose. Nevertheless, many mechanistic models in economics and population biology use game theory as a fundamental ingredient [Rasmusen 1989].

*Network theory*

Just like complexity and game theory, network theory is an excellent place for information exchange between the natural and social sciences. Its basic idea is to use networks of vertices, connected pairwise by edges, to simplify a system systematically. By studying the network structure (roughly speaking, how a network differs from a random network), one can say something about how the system functions as a whole or the roles of the individual vertices and edges in the system [Newman 2010, Barabási 2015]. The multidisciplinarity of network theory is reflected in its overlapping terminology—vertices and edges are called nodes and links in computer science, sites, and bonds in physics and chemistry, actors and ties in sociology, etc.

Many ideas in network theory originated in social science, so it may not fit in a section about influences from natural science. Nevertheless, it is a field where ideas frequently flow from the natural and formal sciences to the social sciences. Centrality measures like PageRank and HITS were, for example, developed in computer science [Newman 2010], as were fundamental concepts of temporal network theory (where information about the time when vertices and edges are active is included in the network) [Holme Saramäki 2012].

## Early computer simulations to understand social mechanisms

In this section, we will go through some developments in using mechanistic models in social science. We will focus on early studies, assuming the readers largely know the current trends. This is by no means a review (which would need volumes of books), but a few snapshots highlighting some differences and similarities to today's science in the methodologies and the questions asked.

*Operations research*

Just like the computer hardware, the research topics for simulation and mechanistic models have many roots in military efforts around the Second World War. Perhaps the main discipline for this type of research in operations research, which is usually classified as a branch of applied mathematics. Operations research aims to optimize the management of large-scale organizations—managing supply chains, scheduling crews of ships, planes, and trains, etc. The military was not the only such organization that interested the early computer simulation researchers. Harling [1958] provides an overview of the state of computer simulations in operation research in the late 1950s. As a typical example, Jennings and Dickins modeled the flow of people and buses in the Port Authority Bus Terminal in New York City during the morning rush hour [Jennings Dick-ins 1958]. They modeled the buses individually and passengers as numbers of exiting, not transferring, individuals. The authors tried to simultaneously optimize the interests of three actors—the bus operators, the passengers, and the Port Authority (operating the terminal). These objectives were mostly not conflicting—in principle, it was better for all if the passenger throughput was as high as possible. A further simplifying factor was that the station was the terminus for all buses. The challenge was that buses stopping to let off passengers could block other buses, thus creating a traffic jam. The paper evaluated different methods to assign a bus stop to an incoming bus to solve this problem.

*Political Science*

Although rarely cited today, simulation studies of political decision processes were quite common in the 1950s and '60s. Crecine [1968] reviews some of these models. One difference from today is that these models were less abstract, often focusing on a particular political or juridical organization. The earliest paper we know about is Guetzkow's 1959 investigation of computer simulations as a support system for international politics [Guetzkow 1959]. However, many studies in this field credit de Sola Pool & al.'s simulation of the American presidential elections 1960 and 1964 as the starting point [de Sola Pool & al. 1965]. In their work, the authors gathered a collection of 480 voter profiles that they could use to test different scenarios (concerning what topics would be necessary for the campaign). Eventually, they predicted the outcome of the elections with 82% accuracy.

In their Ph.D. theses, Cherryholmes [1966] and Shapiro [1966] modeled voting in the House of Representatives by: First, dividing members into classes with respect to how susceptible they were to influence. Second, modeling the influence process via an interaction network where people were more likely to communicate (and thus influence each other) if they were from the same party, state, committee, etc. Cherryholmes and Shapiro also validated their theories against actual voting behavior (something rarely seen in today's simulation studies of opinion spreading [Castellano & al. 2009]). Other authors addressed more theoretical issues of voting systems, such as Arrow's paradox [Klahr 1966, Tullock Campbell 1970] (which states, briefly speaking, that a perfect voting system is impossible for three or more alternatives).

There was also considerable early interest in simulating decision-making within an organization. The Cuban missile crisis of 1962 was an important source of inspiration. De Sola Pool was, once again, a pioneer in this direction with a simulation of decision-making in a developing, general crisis with incomplete information [Kessler De Sola Pool 1965]. Even more explicitly, Smith [1970] based his simulation on the personal accounts of the people involved in solving the Cuban missile crisis. Clema and Kirkham proposed yet another model of risks, costs, and benefits in political conflicts [Clema Kirkham 1971]. Curiously, as late as 2007, a paper was published on simulating the Cuban missile crisis [Stover 2007]. However, this paper explores mechanistic modeling as a method of teaching history rather than the mechanisms of the decision-making process itself.

Another type of political science research concerns the evolution of norms. A classic example is Axelrod's 1986 paper [Axelrod 1986], where he investigated norms emerging as successful strategies in situations described by game theory. Axelrod let the norms evolve by genetic algorithms (an algorithmic framework for optimization inspired by genetics). In addition to norms, Axelrod also studied metanorms—norms that promote other norms (by, e.g., encouraging punishing of people breaking or questioning the norms). Axelrod interpreted the simulation results in terms of established social mechanisms supporting the existence of norms (domination, internalization, deterrence, etc.).

*Linguistics*

In linguistics, the first computer simulation studies appeared in the mid-1960s. A typical early example is Klein [1966], who developed an individual-based simulation platform for the evolution of language. Just like Cherryholmes and Shapiro (above), Klein assumed that the communication was not uniformly random between all pairs of individuals—spouses were more likely to speak to and learn from one another, as were parents and children. In multilingual societies, speakers were more likely to communicate with another speaker of the same language (Klein allowed multilingual individuals). A language was represented by a set of explicit grammatical rules (with explicit word classes: nouns, verbs, etc.). Communication reinforced the grammatical rules between the speakers. Klein incremented the time by years and simulated several generations of speakers but was not entirely happy with the results as communities tended to lose the diversity of their grammar quickly or diverge to mutually incomprehensible grammars. In retrospect, we feel like it was still a significant step forward, where the negative results helped raise essential questions about what mechanisms were missing. More modern models of language evolution have considered much simpler problems [Perfors 2002]. One cannot help thinking that this is to avoid the complexities of reality, and more models in the vein of Klein's 1966 paper would be more important. Later, Klein focused his research on more specific questions like the evolution of Tikopia and Maori [Klein & al. 1969]. The goal of these early simulation studies was to create something similar to a sociolinguistic fieldwork study. Thus these were proof-of-concept studies on a more concrete level than today's more theoretically motivated research.

*Geography*

Demography and geography were also early fields to adopt computer simulations. One notable pioneer was the authors' compatriot Torsten Hägerstrand whose Ph.D. thesis used computer simulations to investigate the diffusion of innovations [Hägerstrand 1953]. His model was similar to two-dimensional disease-spreading models (but probably developed independently of computational epidemiology, where the first paper was published the year before [Abbey 1952]). Hägerstrand used an underlying square grid. People were spread out over the grid according to an empirically measured population distribution. At each simulation iteration, there was a contact between two random individuals (where the chance of contact decayed with their separation). If one of the individuals had adopted the innovation, and the other had not, then the latter would (with 100% probability) adopt it. A goal of Hägerstrand's modeling was to recreate a "nebula shaped" distribution of the innovation (this is further developed in Hägerstrand [1965]). To this end, Hägerstrand introduced a concept (still in use) called "mean information field," representing the probability of getting the information (innovation) from the source.

A technically similar topic to information diffusion is migration (as in moving one's home). This research dates back to Ravenstein's 1885 paper "The laws of migration," which is very mechanistically oriented [Ravenstein 1885]. He listed seven principles for human migration, e.g.: short-distance migration is more common than long-distance; people who migrate far tend to go to a "great centre of commerce or industry." Computer simulation lends itself naturally to exploring the outcomes of such mechanisms in terms of demographics. One such example is Porter's migration model, where agents were driven by the availability of work, and the availability of work was partly driven by where people were. If there were an excess of workers, workers would move to the closest available job opportunity; if there were vacancies, the closest applicant would be offered the job [Porter 1956].

The study of human mobility (how people move around in their everyday lives and extreme situations, such as disasters) is an active field of research. It has even been revitalized lately by the availability of new data sources (see, e.g., Brockmann & al. [2006]). One common type of simulation study involving human mobility data aims at predicting outbreaks of epidemic diseases. To model potentially contagious contacts between people, one can use more or less realism. However, even for the most realistic and detailed simulations, there is a choice of using the empirical data to calibrate a model of human mobility [Eubank & al. 2004] or run the simulation on actual mobility data (perhaps with simulations to fill in missing data) [Balcan & al. 2009].

*Economics and management science*

Many early computational studies in economics used simulation techniques for scenario testing [Cohen 1960, Birchmore 1970]. A typical question at the time was to investigate a company's operations at many levels (overlapping with the operations-research section above). Evidently, the researchers saw a future where every aspect of running a business would be modeled—marketing, human resource development, social interaction within the company, the competition with other firms, adoption of new technologies, etc. To make progress, the authors needed to restrict themselves. Birchmore [1970], for example, focused on forest firms. Much of his work revolved around a forestry firm's interaction with its resource and the many game-theoretical considerations that arose from the conflicting time perspectives of short- and long-time revenues and the com-petition with other companies. Birchmore only used a few combinations of parameter values rather than investigating the parameter dependence like modern game theory would. Finally, we note that economics and management science was also early to address questions about validation and other epistemological aspects of computer simulations [Naylor Finger 1967].

*Anthropology and demographics*

Anthropology was also early to embrace simulation techniques, especially regarding social structure, kinship, and marriage [Coult Randolph 1965]. These are perhaps the traditional problems of anthropology that have the most complex structure of causal explanations, and for that reason, are most in need of proof-of-concept-type computer simulations. Gilbert and Hammel [1966], for example, addressed the question: "How much, and in what ways, is the rate of patrilateral parallel cousin marriage influenced by the number of populations involved in the exchange of women, by their size, by their rules of postmarital residence, and by degree of territorially endogamic preference?" To answer these questions, the authors constructed a complex model including villages of exact sizes, individuals of explicit gender, age and kinship, and rules for selecting a spouse. The model was described primarily in words, much detail, and great length. A modern reader would think that pseudocode would make the paper more readable (and certainly much shorter). The anthropology journals of the time were probably too conservative, or the programming literacy too low, for including pseudocode in the articles.

In a study similar to Gilbert and Hempel, one step closer to demographics, May and Heer [1968] used computer simulations to argue that the large family sizes in rural India (of that time) were rational choices for the individuals than a consequence of ignorance and indecision. Around the same time, there were studies of more general questions of human demographics [Barrett 1969], highlighting a transition from mechanistic models for scenario testing to proof-of-concept models and hypothesis discovery.

*Cognitive and behavioral science*

In cognitive science (sometimes bordering to behavioral science), researchers in the 1960s were excited about the prospects of understanding human cognition as a computer program.

Abelson Carrol [1965], for example, proposed that mechanistic simulations could address questions like how a person can reach an understanding ("develop a belief system") of a complex situation in terms of a set of consistent descriptive clauses (encoding, for example, causal relationships). Several researchers proposed reverse engineering of human thinking into computer programs as a method to understand cognitive processes [Newell Simon 1961]. Some even went so far as to interpret dreams as an operating system process [Newman Evans 1965]. These ideas were not without criticism. Frijda [1967] argued that computer code would always have technical aspects without a corresponding cognitive function. History seems to have given the author right since few studies nowadays pursue replicating human thinking by procedural computer programs. There were, of course, many other types of studies in this area. For example, early studies in computational neuroscience influenced the behavioral-science side of cognitive science [Green 1961].

*Sociology*

Simulation, in sociology, has always been linked to finding social mechanisms. Even before computer simulations, there were mathematical models for that purpose [Edling 2002, Coleman 1964]. As an example of mathematical model building, we briefly mention Nicholas Rashevsky and his program in "mathematical biophysics" at the University of Chicago [Cull 2007, Abraham 2004]. Trained as a physicist, Rashevsky and his group pioneered the modeling of many social (and biological) phenomena such as social influence [Rashevsky 1949], how social group structure affect information flow [Rapoport 1953], and fundamental properties of social networks [Solomonoff Rapoport 1951]. However, Rashevsky and colleagues operated somewhat disconnected from the rest of academia—mostly publishing in their *Bulletin of Mathematical Biophysics* and often not building on empirical results available. Perhaps for this reason (even though his contemporaries were aware of his work [Karlsson 1958]) is Rashevsky & al.'s direct impact on today's sociology is rather limited.

Even though there were stochastic models in sociology in the early 1960s (e.g., White [1962]), these were analyzed analytically, and early sociological computer simulations were off to a relatively late start. Coleman [1965] and Gullahorn and Gullahorn [1963], and [1965] gave the earliest discussions of the prospects of computer modeling in sociology that we are aware of. Coleman discussed abstract questions about social action and social organization and more concrete ones like using simulation to test social-contagion scenarios of smoking among adolescents. The Gullahorns were more interested in organization and conflict resolution, typically in the interface of sociology and behavioral science. McGinnis [1968] presented a stochastic social mobility model that he analyzed analytically and by simulations. "Mobility," in McGinnis work, should be understood in an extremely general sense, indicating a change of an individual's position in any sociometric observable (including physical space).

Markley's 1967 paper on the SIVA model is another early simulation study of a classic sociological problem [Markley 1967], namely what kind of pairwise relationships could build up a stable organization. The letters SIVA stands for four aspects of such relationships in an organization facing some situation that could require some action to be taken—Strength (the ratio of how vital the two individuals are to the organization), influence (describing how strongly they influence each other), volitional (the relative will to act depending on the situation) and action (quantifying the joint result of the two actors). These different aspects are coupled, and Markley used computer simulations to find fixed points of the dynamics. For many parameter values, it turned out that the SIVA values diverged or fluctuated—which Markley took as an indication that one would not observe such combinations of parameter values in real organizations.

A model touching classical sociological ground that recently has received exceptional amounts of attention is Schelling's segregation model [Schelling 1969]. With this model, Schelling argued that strong racial segregation (with the United States in mind) does not necessarily mean that people have firm desires about the race of their neighbors. Briefly, Schelling spread individuals of two races on a square grid. Some sites were left vacant. Then he picked an individual at random. If this individual had a lower ratio of neighbors of the same race than a threshold value, then he or she moved to a vacant site. It turned out that the segregation (measured as the fraction of links between people of the same race) would always move away from the threshold as the iterations converged. Segregation, Schelling concluded, could thus occur without people actively avoiding different races (they just needed to seek similar neighbors), and spatial effects would make a naïve interpretation of the observed mixing overestimating the actual sentiments of the people. The core question—what are the weakest requirements (of tolerance to your neighbors' ethnicity) for something (racial segregation) to happen—was a hallmark of Schelling's research and probably an approach that could be fruitful for future studies. We highly recommend Schelling's popular science book *Micromotives and Macrobehavior* [Schelling 1978] as a bridge between the methodologies of natural and social science.

## Discussion and conclusions

The motivation for using mechanistic models in social science is often as proof-of-concept models. "[I]t forces one to be specific about the variables in interpersonal behavior and the exact relation between them" [Hare 1961, Gullahorn Gullahorn 1963, Hartman Walsh 1969]. Computer programming forces researchers to break down the social phenomena into algorithmic blocks to help identify mechanisms [Dutton Briggs 1971, Gullahorn Gullahorn 1963]. Other authors point out that with computational methods, the researchers can avoid oversimplifying the problem [De Sola Pool & al. 1965]. Another point of view is that simulation in social sciences is primarily for exploring poorly understood situations and phenomena as a replacement for an actual (in practice impossible to carry out) experiment [Simon 1969, Fleisher 1965, Naylor & al. 1969, Crecine 1968]. Such models are closest to hypothesis generators in our above classification. Crane [1962] and Ostrom [1988] think of computer simulations as languages for social science, alongside natural languages and mathematics. Going a bit off-topic, other authors went so far as to using or recommending using computer programs as representations of human cognitive processes [Colby 1967, Newell Simon 1961, Newman Evans 1965].

As illustrated by our examples, the history of computational studies in social science has seen a gradual shift of focus. In the early days, it was, as mentioned, often regarded as a replacement for empirical studies. Such mechanistic models for scenario testing still exist in both natural and social science. However, nowadays, it is much more com-mon to use computational methods in theory building—either one uses it to test the completeness of a theoretical framework (proof-of-concept modeling), or to explore the space of possible mechanisms or outcomes (hypothesis discovery).

It is quite remarkable how similar this development has been in the natural and social sciences. At least since the mid-1950s, it is hard to say that one side leads the way. This is reflected in how the information flows between disciplines. Looking at the interdisciplinary citation patterns [Rosvall Bergstrom 2011] found that out of 203,900 citations from social science journals, 33,891 were to natural science journals, and out of 10,080,078 citations from natural science journals, 35,199 were to social science journals. If citations were random, without any within-field bias, there would be around 201,000 interdisciplinary citations in both directions, 5.9 times the number of social science citations to natural science and 5.7 times the number of natural science citations to social science. In this view, there is almost no inherent asymmetry in the information flow between the areas, only an asymmetry induced by the size difference.

Even though social scientists do not need to collaborate with natural scientists to develop mechanistic modeling, we encourage collaboration. The usefulness of interdisciplinary collaborations comes from the details of the scientific work. It can help people to see their object system with new eyes. One discipline may, for example, care about the extreme and need input from another to see exciting aspects of the average (cf. phase transitions in the complexity of algorithms [Moore Mertens 2011]). Interdisciplinary information flow could help a discipline overcome technical difficulties. The use of MCMC techniques in the social sciences may be a good example of this. It is, however, important that such developments come from a need to understand the world around us and not just because they have not been done before.

A major trend at the time of writing is "big data" and "data science". This essay has intentionally focused on the other side of computational social science—mechanistic models. In practice, these two sides can (and do) influence each other. If it cannot predict real systems at all, a mechanistic model is useless in providing a causal explanation [Watts 2014, Hindman 2015]. Modern, large-scale data sets provide plenty of opportunities to validate models [Lazer & al. 2009, Holme Huss 2011, Pentland 2014]. Another use of big data is in hybrid approaches where one combines a simulation and an empirical dataset, for example, simulations of disease spreading on temporal networks of human contacts [Holme Saramäki 2012].

As a concluding remark, we want to express our support for social scientists interested in exploring the methods of natural science and natural scientists seeking applications for their methods in the social sciences. To be successful and make the most out of such a step, we recommend that the social scientist spend a month learning a general programming language (Python, Matlab, C, etc.). There is no shortcut (like an integrated modeling environment) to learning the computational subtleties and tradeoffs of building a simulation model, and simulation papers often do not mention them. Furthermore, if a social scientist leaves this aspect to a natural scientist, she also leaves parts of the social modeling to the natural scientist—collaboration simply works better if the computational fundamentals need not be discussed. To the theoretical natural scientists that are used to simulations, we recommend spending a month reading popular social science books (e.g. [Watts 2012, Simon 1969, Schelling 1978]). There are too many examples of natural scientists going into social science with the ambition to use the same methods as they are used to—only replacing the natural components by social—and ending up with unverifiable results, too general to be informative, infeasible, or already known. While reading, we encourage meditating on the following question: Why do social scientists ask different questions about society than natural scientists about nature?

## Acknowledgments

The authors thank Martin Rosvall for the citation statistics.

Funding: This research was supported by the Basic Science Research Program through the National Research Foundation of Korea (NRF) funded by the Ministry of Education (2013R1A1A2011947).